\documentclass[times,sort&compress,3p]{elsarticle}

\usepackage{float,amssymb,amsmath,amsthm,bbm,rotating,supertabular,placeins}
\usepackage[pdfpagelabels]{hyperref}

\journal{arXiv}

\begin{document}

\begin{frontmatter}

\title{Metric projection for dynamic multiplex networks}

\author{Giuseppe Jurman}
\address{Fondazione Bruno Kessler, Trento, Italy}
\ead{jurman@fbk.eu}

\begin{abstract}
Evolving multiplex networks are a powerful model for representing the dynamics along time of different phenomena, such as social networks, power grids, biological pathways.
However, exploring the structure of the multiplex network time series is still an open problem.
Here we propose a two-steps strategy to tackle this problem based on the concept of distance (metric) between networks.
Given a multiplex graph, first a network of networks is built for each time steps, and then a real valued time series is obtained by the sequence of (simple) networks by evaluating the distance from the first element of the series.
The effectiveness of this approach in detecting the occurring changes along the original time series is shown on a synthetic example first, and then on the Gulf dataset of political events.
\end{abstract}


\end{frontmatter}

\section{Introduction}
\label{sec:intro}
When the links connecting a set of $N$ nodes arise from $k$ different sources, a possible representation for the corresponding graph is the construction of $k$ networks on the same $N$ nodes, one for each sources.
The resulting structure is known as a multiplex networks, and each of the composing graphs is called a layer.
Multiplex networks are quite effective in representing many different real-world situations\cite{boccaletti14structure,dedomenico13mathematical,kivela14multilayer}, and their structure helps extracting crucial information about the complex systems under investigation that would instead remain hidden when analysing individual layers separately~\cite{menichetti14weighted,weiyi15mining}; furthermore, their relation with time series analysis techniques has recently gained interest in the literature~\cite{lacasa15network}.
A key property to be highlighted is the correlated multiplexity, as stated in~\cite{lee15towards}: in real-world systems, the relation between layers is not at all random; in fact, in many cases, the layers are mutually correlated.
Moreover, the communities induced on different layers tend to overlap across layers, thus generating interesting mesoscale structures.

These observations guided the authors of~\cite{iacovacci15mesoscopic} in defining a network having the layers of the original multiplex graph as nodes, and using information theory to define a similarity measure between the layers themselves, so to investigate the mesoscopic modularity of the multiplex network.
Here we propose to pursue a similar strategy for definining a network of networks derived from a multiplex graph, although in a different context and with a different aim.
In particular, we project a time series of multiplex networks into a series of simple networks to be used in the analysis of the dynamics of the original multiplex series.
The projection map defining the similarity measure between layers is induced by the HIM network distance~\cite{jurman15him}, a glocal metric combining the Hamming and the Ipsen-Mikhailov distances, used in different scientific areas~\cite{furlanello13sparse,mina15tumor,fay15graph,gobbi15null,masecchia13statistical,csermely13structure}.
The main goal in using this representation is the analysis of the dynamics of the original time series through the investigation of the trend of the projected evolving networks, by extracting the corresponding real-valued time series obtained computing the HIM distance between any element in the series and the first one.

For instance, we show on a synthetic example that this strategy is more informative than considering statistics of the time series for each layer of the multiplex networks, or than studying the networks derived collapsing all layers into one including all links, as in~\cite{dedomenico15structural,cardillo13emergence} when the aim is detecting the timesteps where more relevant changes occur and the system is undergoing a state transition (tipping point) or it is approaching it (early warning signals).  
This is a classical problem in time series analysis, and very diverse solutions have appeared in literature (see~\cite{takenaka15detecting} for a recent example).
Here we use two different evaluating strategies, the former based on the fluctuations of mean and variance~\cite{killick14changepoint} (implemented in the R package \textit{changepoint} \url{https://cran.r-project.org/web/packages/changepoint/index.html}), and the latter involving the study of increment entropy indicator~\cite{liu15increment}. 

We conclude with the analysis of the well known Gulf Dataset (part of the Penn State Event Data) concerning the 304.401 political events (of 66 different categories) occurring between 202 countries in the 10 years between 15 April 1979 to 31 March 1999, focussing on the situation in the Gulf region and the Arabian peninsula. 
A major task in the analysis of the Gulf dataset is the assessment of the translation of the geopolitical events into fluctuations of measurable indicators.
A similar network-based mining of sociopolitical relations, but with a probabilistic approach, can be found in~\cite{schein13inferring,schein15bayesian,jackson15networks}.
Here we show the effectiveness of the newly introduced methodology in associating relevant political events and periods to characteristic behaviours in the dynamics of the time series of the induced networks of networks, together with a simple overview of the corresponding mesoscale modular structure.

\section{Theory}
\label{sec:methods}
\begin{figure}[!ht]
\includegraphics[width=0.95\textwidth]{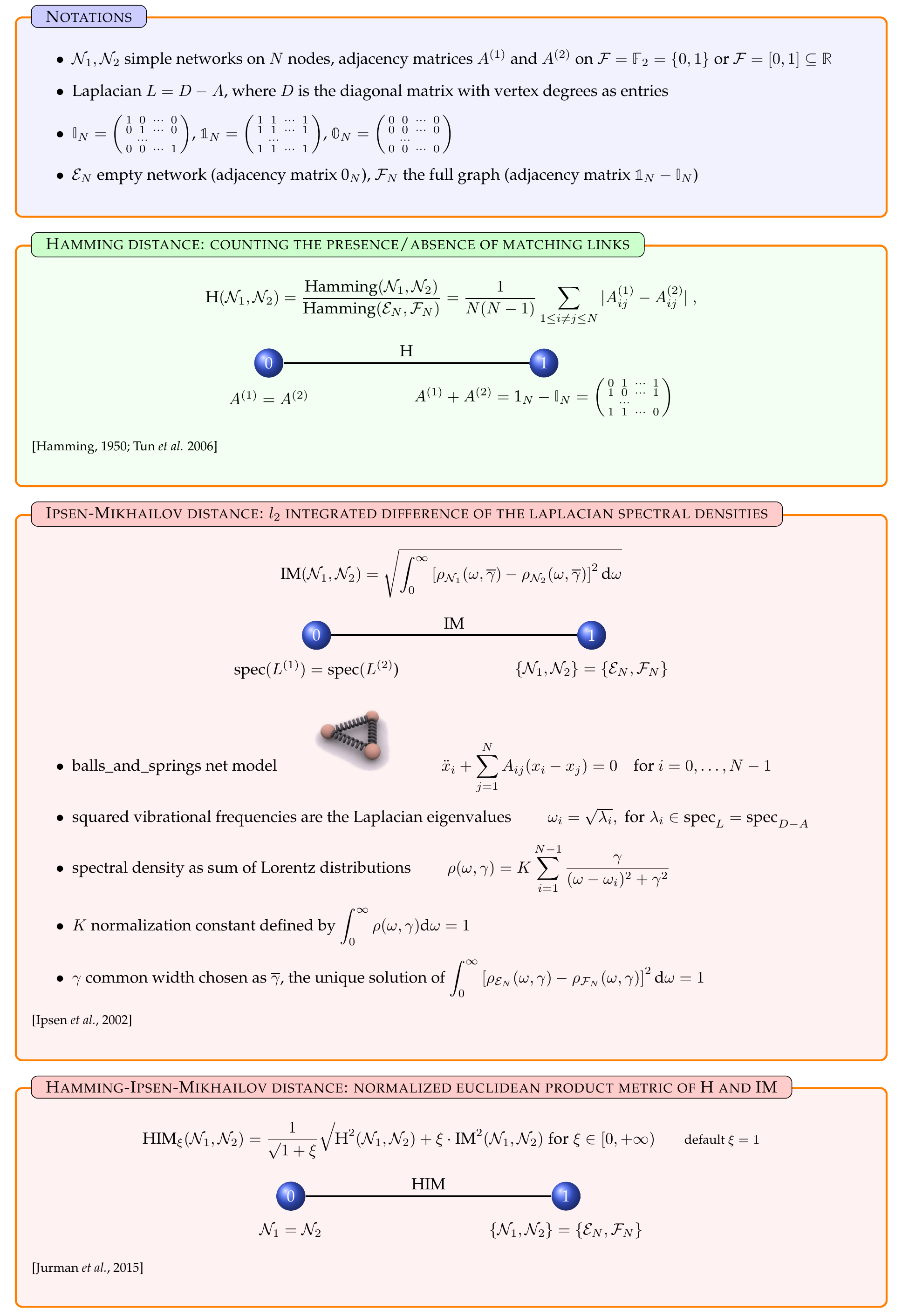}
\caption{Summary of the definitions of the HIM distance and its Hamming (H) and Ipsen-Mikhailov (IM) components.}
\label{fig:him}
\end{figure}

Let $\mathcal{N}=\{\mathcal{N}(t)\}_{t=1}^\tau$ be a sequence (time series) of $\tau$ multiplex networks with $\lambda$ layers $\{L_i(t)\}_{i=1}^\lambda$ sharing $\nu$ nodes $\{v_j\}_{j=1}^\nu$.
Construct now the metric projection $\mathcal{LN}(t)$ of $\mathcal{N}(t)$ as the full undirected weighted network with $\lambda$ nodes $\{w_{L_i}\}_{i=1}^\lambda$ where the weight of the edge connecting vertices $w_{L_i}$ and $w_{L_j}$ is defined by the HIM similarity between layers $L_i(t)$ and $L_j(t)$: thus, if $A^\mathcal{LN}(t)$ is the adjacency matrix of $\mathcal{LN}(t)$, then 
\begin{displaymath}
A_{ij}^\mathcal{LN}(t)  = 1-\textrm{HIM}(L_i(t),L_j(t))\ .
\end{displaymath}
See Fig.~\ref{fig:him} for a quick recap on the HIM distance: further details in~\cite{jurman15him,jurman14him}.
Moreover, if $A^{L_i}(t)$ is the adjacency matrix of $L_i(t)$, define the collapsed projection $\mathcal{CN}(t)$ of $\mathcal{N}(t)$ on nodes $\{v_j\}_{j=1}^\nu$ as the network where a link exists between $v_k$ and $v_q$ if it exists in at least one layer $\{L_i(t)\}_{i=1}^\lambda$ (for binary layers); in case of weighted layers, the weight of the link $v_k - v_q$ is the average of the weights across all layers.
Thus, if  $A^\mathcal{CN}(t)$ is the adjacency matrix of $\mathcal{CN}(t)$, then 
\begin{displaymath}
A_{kq}^\mathcal{CN}(t)  = 
\begin{cases}
\displaystyle{\bigvee_{i=1}^\lambda} A_{kq}^{L_i}(t) & \textrm{for binary layers}\\
\\
\displaystyle{\frac{1}{\lambda} \sum_{i=1}^\lambda} A_{kq}^{L_i}(t) & \textrm{for weighted layers} \ .
\end{cases}
\end{displaymath}
Caveat: consider a sequence of binary multiplex networks such that, for each of the possible $\frac{\nu(\nu-1)}{2}$ links and for each timestep, there exists at least one layer including this link. 
Then the collapsed projection, at each time step, is the full graph on $\nu$ nodes, and, as such, it has no temporal dynamics, regardless of the evolution of each single layer.

To investigate the dynamics of $\mathcal{N}(t)$ for $t=1,\ldots,\tau$, we construct a suite of associated time series by means of three different procedures, all involving the HIM distance between each network in a given sequence and the first element of the sequence itself.
The first group D1 of distance series is obtained by evaluating the dynamics of each layer considered separately:
\begin{displaymath}
\left\{ \textrm{HIM}(L_i(t),L_1(t)),\; t=2,\ldots,\tau\right\}\quad i=1,\ldots,\lambda\ .
\tag{D1}
\end{displaymath}
The second series, D2, collects the metric dynamics of the collapsed projection ${\mathcal{CN}}$:
\begin{displaymath}
\textrm{HIM}({\mathcal{CN}}(t),{\mathcal{CN}}(1)),\; t=2,\ldots,\tau\ .
\tag{D2}
\end{displaymath}
Finally, the last series D3 collects the metric dynamics of the metric projection $\mathcal{LN}$:
\begin{displaymath}
\textrm{HIM}(\mathcal{LN}(t),\mathcal{LN}(1)),\; t=2,\ldots,\tau\ .
\tag{D3}
\end{displaymath}
\FloatBarrier

\section{Results and discussion}
\subsection{A synthetic example}
\label{sec:synth}
Consider now a sequence of binary multiplex networks with $\tau=30$, $\lambda=5$ and $\nu=10$, generated as follows.

Define the perturbation function $\Pi(N,(m,M))$ taking as entries a binary simple network $N$ on $n$ nodes, and a couple of real values $(m,M)$ with $0\leq m\leq M\leq 1$, and returning a network $N^\prime$ obtained from $N$ by swapping the status (present/not present) of $\lfloor g\frac{n(n-1)}{2}\rfloor$ links, where $g$ is a random value in the interval $[m,M]$.
Further, define the default transition as the pair $\sigma_d=(0.05,0.2)$, a small transition as $\sigma_s=(0.2,0.3)$, a medium transition as $\sigma_m=(0.25,0.4)$ and, finally, a large transition as $\sigma_l=(0.5,0.7)$.
Moreover, let $R$ be an Erd{\'o}s-R{\'e}nyi $G(\nu,0.3)$ random model and define 4 special timepoints: the initial time step $\tau_0=1$, the first spike $\tau_1=10$, the second spike $\tau_2=17$ and the third spike $\tau_3=24$.

\begin{figure}[!t]
\includegraphics[width=0.95\textwidth]{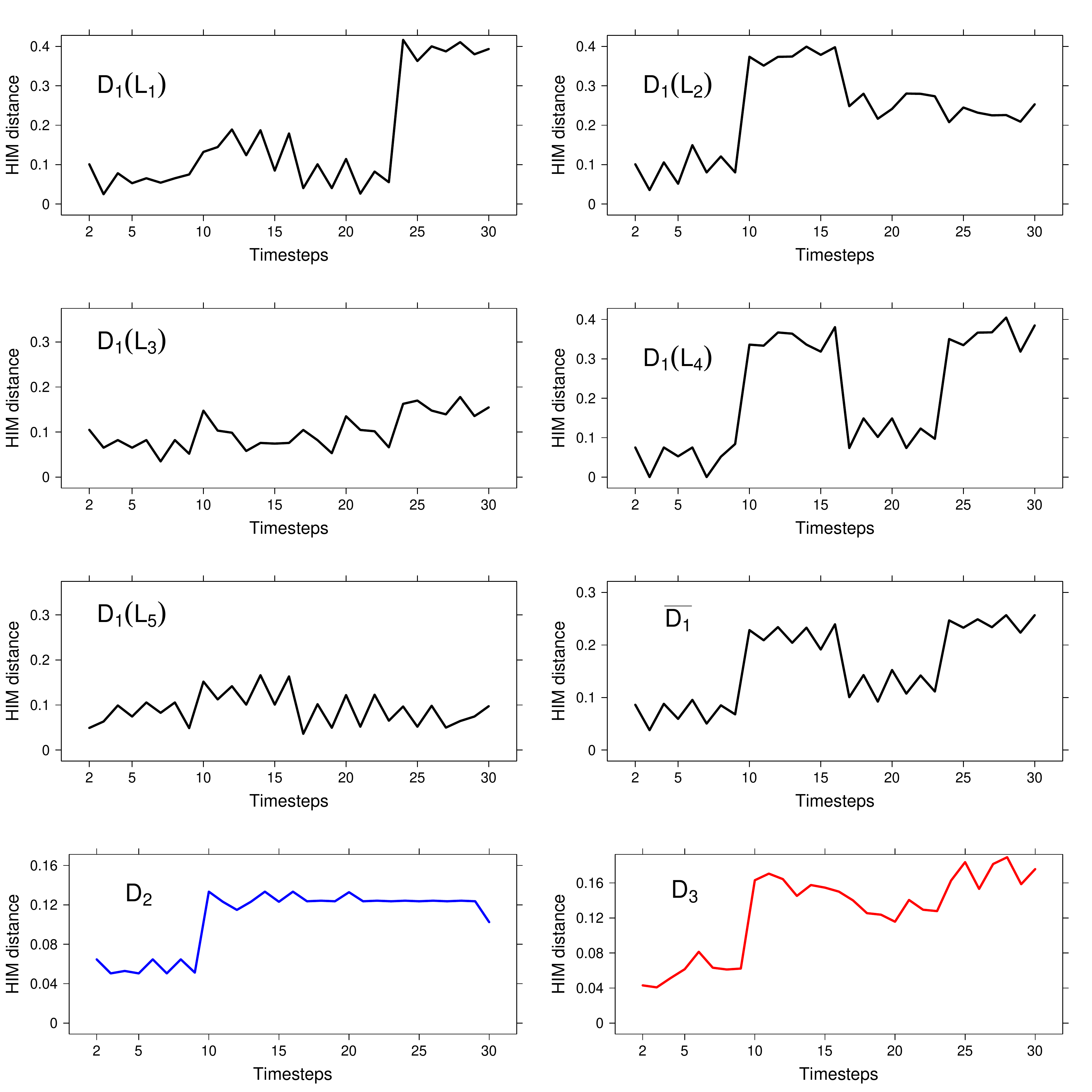}
\caption{$D_1$, $D_2$, $D_3$ for a synthetic example on 5 layers and 30 timepoints; in the right column, third row, we plot $\overline{D}_1=\frac{1}{5}\sum_{i=1}^5 D_i(L_i)$.}
\label{fig:synth}
\end{figure}

Then, each layer $L_i$ at a given time step $t$ is defined through the following rule:
\begin{displaymath}
L^i(t)=
\begin{cases}
\Pi(R,\sigma_s) & \textrm{if $t=\tau_0$ and $i=1,2$}\\
\Pi(R,\sigma_m) & \textrm{if $t=\tau_0$ and $i=3,4$}\\
\Pi(R,\sigma_l) & \textrm{if $t=\tau_0$ and $i=5$}\\
\Pi(L^i(t-1),\sigma_s) & \textrm{if $t=\tau_1$ and $i=1,3,5$}\\
                       & \textrm{or if $t=\tau_2$ and $i=3,5$}\\
                       & \textrm{or if $t=\tau_3$ and $i=5$}\\
\Pi(L^i(t-1),\sigma_m) & \textrm{if $t=\tau_2$ and $i=1,2$}\\
                       & \textrm{or if $t=\tau_3$ and $i=3$}\\
\Pi(L^i(t-1),\sigma_l) & \textrm{if $t=\tau_1$ and $i=2,4$}\\
                       & \textrm{or if $t=\tau_2$ and $i=4$}\\
                       & \textrm{or if $t=\tau_3$ and $i=1,2,4$}\\
\Pi(L^i(t-1),\sigma_d) & \textrm{otherwise}\ .
\end{cases}
\end{displaymath}
In Fig.~\ref{fig:synth} we show the evolution along the 30 timepoints of the 5 curves for $D_1(L_i)$, its average $\overline{D}_1=\frac{1}{5}\sum_{i=1}^5 D_i(L_i)$ and $D_2$, $D_3$.
To assess the information content of each curve we adopt the Increment Entropy (IncEnt) indicator~\cite{liu15increment}, whose value increases with the series' complexity: the IncEnt values are reported in Tab.~\ref{tab:incent}.
\begin{table}[!b]
\begin{center}
\caption{Increment Entropy values for the distance sequences of the synthetic example, with parameters $m=2$, $R=2$.}
\label{tab:incent}
\begin{tabular}{lr|lr}
Dist. & IncEnt & Dist. & IncEnt \\
\hline
$D_1(L_1)$ & 2.52 & $D_1(L_2)$ & 2.78 \\
$D_1(L_3)$ & 2.59 & $D_1(L_4)$ & 2.83 \\
$D_1(L_5)$ & 2.44 & $\overline{D}_1$ & 2.27 \\
$D_2$ & 1.82 & $D_3$ & 3.04 
\end{tabular}
\end{center}
\end{table}

Among the evolving layers, $L_2$ and $L_4$ have the largest IncEnt, while the other three layers show a lower level of complexity. 
As expected, the average $\overline{D_1}$ and the collapsed network distance $D_2$ has very low IncEnt value, yielding that both averaging the distances and collapsing the layers lose information about the overall dynamics.
Finally, distance $D_3$ is the metric which better detects the network evolution along time, conserving most of the information.
This is also supported by the changepoints detection indicator mean/variance (meanvar)~\cite{killick14changepoint,chen12parametric,scott74cluster,auger89algorithms,zhang07modified}, with the CROPS penalty $[2\log(\tau),10\log(\tau)]$~\cite{haynes14efficient} in the PELT method~\cite{killick12optimal}. 
In fact, the meanvar indicator detects correctly in $D_3$ the three points $\tau_1-1$, $\tau_2-1$ and $\tau_3-1$ as changepoints, while in $D_2$, other than the $\tau_1-1$, meanvar detects $t=20$ and $=28$ which are unrelated to the designed dynamics.
\FloatBarrier

\subsection{The Gulf Dataset} 
\label{sec:gulf}

\paragraph{Data description}
\label{ssec:data}
Part of the Penn State Event Data \url{http://eventdata.psu.edu/} (formerly Kansas Event Data System), available at \url{http://vlado.fmf.uni-lj.si/pub/networks/data/KEDS/}, the Gulf Dataset collects, on a monthly bases, political events between pairs of countries focusing on the Gulf region and the Arabian peninsula for the period 15 April 1979 to 31 March 1999, for a total of 240 months. 
The 304401 political events involve 202 countries and they belong to 66 classes (including for instance ''pessimist comment'', ''meet'', ''formal protest'', ''military engagement'', etc.) as coded by the World Event/Interaction Survey (WEIS) Project~\cite{mcclelland76world,goldstein04international,goldstein92conflict} \url{http://www.icpsr.umich.edu/icpsrweb/ICPSR/studies/5211}, whose full list is reported in Tab.~\ref{tab:weis1},\ref{tab:weis2}.

\begin{table}[!b]
\begin{center}
\caption{Part 1 of the full table of WEIS codes~\cite{mcclelland76world}, with the 66 layers considered in the Gulf dataset case study; entries with no layer number were not monitored in the Gulf dataset events collection.}
\label{tab:weis1}
\begin{tabular}{rrcl}
Layer\# & WEIS code & WEIS cat & Description \\
\hline
1 & 011 & Yield & Surrender, yield or order, submit to arrest, etc.\\ 
2 & 012 & Yield & Yield position, retreat; evacuate \\
3 & 013 & Yield & Admit wrongdoing; retract statement \\
  & 014 & Yield & Accommodate, Cease-fire \\
4 & 015 & Yield & Cede Power \\
5 & 021 & Comment & Explicit decline to comment  \\
6 & 022 & Comment & Comment on situation -- pessimistic \\
7 & 023 & Comment & Comment on situation -- neutral\\
8 & 024 & Comment & Comment on situation -- optimistic\\
9 & 025 & Comment & Explain policy or future position \\
  & 026 & Comment & Appoint or Elect\\
  & 027 & Comment & Alter Rules\\
10 & 031 & Consult & Meet with at neutral site, or send note.\\
11 & 032 & Consult & Consult \& Visit; go to\\
12 & 033 & Consult & Receive visit; host\\
   & 034 & Consult & Vote, Elect\\
13 & 041 & Approve & Praise, hail, applaud, condole\\
14 & 042 & Approve & Endorse other's policy or position; give verbal support\\
   & 043 & Approve & Rally\\
15 & 051 & Promise & Promise own policy support\\
16 & 052 & Promise & Promise material support\\ 
17 & 053 & Promise & Promise other future support action\\ 
18 & 054 & Promise & Assure; reassure\\
   & 055 & Promise & Promise Rights\\
19 & 061 & Grant & Express regret; apologize \\
20 & 062 & Grant & Give state invitation \\
21 & 063 & Grant & Grant asylum \\
22 & 064 & Grant & Grant privilege, diplomatic recognition \\
23 & 065 & Grant & Suspend negative sanctions; truce \\
24 & 066 & Grant & Release and/or return persons or property \\
   & 067 & Grant & Grant Position \\
25 & 070 & Reward & Reward \\
26 & 071 & Reward & Extend economic aid (as gift and/or loan)\\
27 & 072 & Reward & Extend military assistance \\
28 & 073 & Reward & Give other assistance \\
29 & 081 & Agree & Make substantive agreement \\
30 & 082 & Agree & Agree to future action or procedure; agree to meet, to negotiate\\
   & 083 & Agree & Ally\\
   & 084 & Agree & Merge; Integrate\\
31 & 091 & Request & Ask for information\\
32 & 092 & Request & Ask for policy assistance \\
33 & 093 & Request & Ask for material assistance \\
34 & 094 & Request & Request action; call for \\
35 & 095 & Request & Entreat; plead; appeal to  \\
   & 096 & Request & Request policy change \\
   & 097 & Request & Request rights \\
\hline
\end{tabular}
\end{center}
\end{table}

\begin{table}[!b]
\begin{center}
\caption{Part 2 of the full table of WEIS codes~\cite{mcclelland76world}, with the 66 layers considered in the Gulf dataset case study; entries with no layer number were not monitored in the Gulf dataset events collection.}
\label{tab:weis2}
\begin{tabular}{rrcl}
Layer\# & WEIS code & WEIS cat & Description \\
\hline
36 & 101 & Propose & Offer proposal\\
37 & 102 & Propose & Urge or suggest action or policy\\
38 & 111 & Reject & Turn down proposal; reject protest demand, threat, etc\\
39 & 112 & Reject & Refuse; oppose; refuse to allow\\
   & 113 & Reject & Defy law\\
40 & 121 & Accuse & Charge; criticize; blame; disapprove\\
41 & 122 & Accuse & Denounce; denigrate; abuse\\
   & 123 & Accuse & Investigate\\
42 & 131 & Protest & Make complaint (not formal) \\ 
43 & 132 & Protest & Make formal complaint or protest \\ 
   & 133 & Protest & Symbolic act \\
44 & 141 & Deny & Deny an accusation\\
45 & 142 & Deny & Deny an attributed policy, action role or position\\
46 & 150 & Demand & Issue order or command; insist; demand compliance; etc\\
   & 151 & Demand & Issue Command\\
   & 152 & Demand & Claim Rights\\
47 & 160 & Warn & Give warning\\
48 & 161 & Warn & Warn of policies\\
   & 162 & Warn & Warn of problem\\
49 & 171 & Threaten & Threat without specific negative sanctions  \\
50 & 172 & Threaten & Threat with specific non-military negative sanctions \\
51 & 173 & Threaten & Threat with force specified \\
52 & 174 & Threaten & Ultimatum; threat with negative sanctions and time limit specified \\
53 & 181 & Demonstrate & Non-military demonstration; to walk out on\\
54 & 182 & Demonstrate & Armed force mobilization\\
55 & 191 & Reduce Relations$^1$ & Cancel or postpone planned event \\
56 & 192 & Reduce Relations$^1$ & Reduce routine international activity; recall officials; etc\\
57 & 193 & Reduce Relations$^1$ & Reduce or halt aid \\
58 & 194 & Reduce Relations$^1$ & Halt negotiations \\
59 & 195 & Reduce Relations$^1$ & Break diplomatic relations\\
   & 196 & Reduce Relations$^1$ & Strike\\
   & 197 & Reduce Relations$^1$ & Censor\\
60 & 201 & Expel & Order personnel out of country\\
61 & 202 & Expel & Expel organization or group\\
   & 203 & Expel & Ban Organization\\
62 & 211 & Seize & Seize position or possessions \\
63 & 212 & Seize & Detain or arrest person(s) \\
   & 213 & Seize & Hijack; Kidnap\\
64 & 221 & Force & Non-injury obstructive act \\
65 & 222 & Force & Non-military injury-destruction\\ 
66 & 223 & Force & Military engagement \\
\hline
\end{tabular}
\end{center}
{\small\sl ${}^1$ as negative sanctions}
\end{table}

In the notation of Sec.~\ref{sec:methods}, the Gulf Dataset translates into a time series of $\tau=240$ multiplex networks with $\lambda=66$ unweighted and undirected layers sharing $\nu=202$ nodes.
The landmark event for the considered zone in the 20 years data range of interest is definitely the First Gulf War (FGW), occurring between August 1990 and March 1999. 
However, other (smaller) events located in the area had a relevant impact on world politics and diplomatic relations.
Among them, the Iraq Disarmament Crisis (IDC) in February 1998 significantly emerges from the data, as shown in what follows. 
During that month, Iraq President Saddam Hussein negotiated a deal with U.N. Secretary General Kofi Annan, allowing weapons inspectors to return to Baghdad, preventing military action by the United States and Britain.
\FloatBarrier

\subsection{Network statistics}
\label{ssec:stats}
Consider in this section the set of 304401 edges connecting the 202 nodes independently of their class. 
In Tab.~\ref{tab:nodes} we list the top-10 countries/institutions participating in the largest number of edges across different time spans, together with the absolute number of shared edges and the corresponding percentage over the total number of edges for the period. 
In general, USA, Iraq and Iran are the major players, with different proportions according to the specific period: in particular, Iraq is the main character in both the major events, FGW and IDC.
Other key actors are Israel, the United Nations and the Saudi Arabia, with a relevant presence in each key event in the area.
Note that, overall, the top 20 institutions (also including, other than those listed in the table, the Arab world, France, Syria, Egypt, Russia, Turkey, Jordan, Libya, Germany and the Kurd world) are responsible for 82.57\% of all edges.

\begin{table}[!b]
\begin{center}
\caption{Top-10 countries/institutions ranked by number of shared links, absolute and in percentage over (twice) the total number of links in the considered period. The Iran-Iraq War started in September 1980 and ended in August 1988. SA: Saudi Arabia; UN: United Nations.}
\label{tab:nodes}
\begin{tabular}{crr|crr|crr}
\multicolumn{3}{c|}{Apr79-Mar99} & \multicolumn{3}{c|}{FGW} & \multicolumn{3}{c}{IDC} \\
\hline
\multicolumn{3}{c|}{Edges 304401} & \multicolumn{3}{c|}{Edges 41181}  & \multicolumn{3}{c}{Edges 7712} \\
\hline
Inst. & Degree & \% & Inst. & Degree & \% & Inst. & Degree & \% \\
\hline
USA&93900&15.42&Iraq&18691&22.69&Iraq&3830&24.83 \\
Iraq&84974&13.96&USA&15584&18.92&USA&2876&18.65 \\
Iran&61782&10.15&Kuwait&5245&6.37&UN&1946&12.62 \\
Israel&32204&5.29&SA&3548&4.31&Russia&896&5.81 \\
UN&30097&4.94&Israel&3420&4.15&UK&715&4.64 \\
SA&20503&3.37&UN&3363&4.08&France&651&4.22 \\
Lebanon&19130&3.14&UK&2997&3.64&Iran&468&3.03 \\
Palestine&18607&3.06&Iran&2104&2.55&Arab world&321&2.08 \\
UK&18415&3.02&France&2076&2.52&China&309&2.00 \\
Kuwait&17405&2.86&Arab world&2053&2.49&Kuwait&306&1.98 \\
\hline
\hline
\multicolumn{3}{c|}{Apr79-FGW} & \multicolumn{3}{c|}{FGW-Mar99} &  \multicolumn{3}{c}{Iran-Iraq War} \\
\hline
\multicolumn{3}{c|}{Edges 130990} & \multicolumn{3}{c|}{Edges 132230} & \multicolumn{3}{c}{Edges 95189} \\
\hline
Inst. & Degree & \% & Inst. & Degree & \% & Inst. & Degree & \% \\
\hline
Iran&43818&16.73&USA&43606&16.49&Iran&32812&17.24\\
USA&34710&13.25&Iraq&40677&15.38&USA&24111&12.66\\
Iraq&25606&9.77&UN&19858&7.51&Iraq&21019&11.04\\
Israel&12731&4.86&Israel&16053&6.07&Israel&9189&4.83\\
Palestine&10622&4.05&Iran&15860&6.00&SA&8089&4.25\\
Lebanon&10374&3.96&UK&9209&3.48&Palestine&7521&3.95\\
SA&10290&3.93&Lebanon&8143&3.08&Lebanon&6992&3.67\\
Arab world&8237&3.14&France&6925&2.62&Syria&6072&3.19\\
Syria&8089&3.09&Russia&6875&2.60&Arab world&5726&3.01\\
UN&6876&2.62&SA&6665&2.52&Kuwait&4890&2.57\\
\end{tabular}
\end{center}
\end{table}

Out of all potential $\frac{202\cdot 201}{2}=20301$ unique edges, only $4394$ are represented in the Gulf Dataset. 
In Tab.~\ref{tab:edges} we list the top-10 links ranked by occurrence, together with the number of occurrences itself and the corresponding percentage over the total number of edges for the period.
As it happens for the nodes, there are a few key links throughout the whole timespan which are consistently present in most of the important events, with different proportions.
However, in some of the events, there is an interesting wide gap in the number of occurrences between the very top edges and the remaining ones, \textit{e.g.}, Iraq-USA in FGW (and post) and IDC, and Iran-Iraq during the corresponding war and in the pre-FGW, yielding that these are the links mainly driving the whole network evolution.

\begin{table}[!b]
\begin{center}
\caption{Top-10 countries/institutions ranked by number of shared links, absolute and in percentage over (twice) the total number of links in the considered period. The Iran-Iraq War started in September 1980 and ended in August 1988. SA: Saudi Arabia; UN: United Nations.}
\label{tab:edges}
\begin{tabular}{crr|crr|crr}
\multicolumn{3}{c|}{Apr79-Mar99} & \multicolumn{3}{c|}{FGW} & \multicolumn{3}{c}{IDC} \\
\hline
\multicolumn{3}{c|}{Edges 304401} & \multicolumn{3}{c|}{Edges 41181}  & \multicolumn{3}{c}{Edges 7712} \\
\hline
Edge  & Degree & \% & Edge & Degree & \% &Edge & Degree & \% \\
\hline
Iran-Iraq&19121&6.28&Iraq-USA&6061&14.72&Iraq-USA&1021&13.24\\
Iraq-USA&19002&6.24&Iraq-Kuwait&2306&5.60&Iraq-UN&927&12.02\\
Iran-USA&14051&4.62&SA-USA&1169&2.84&UN-USA&337&4.37\\
Iraq-UN&12775&4.20&Iraq-UN&1118&2.71&Iraq-Russia&315&4.08\\
Israel-Lebanon&6590&2.16&Kuwait-USA&1050&2.55&Iraq-UK&241&3.12\\
Israel-USA&5803&1.91&Iraq-UK&1012&2.46&France-Iraq&191&2.48\\
Iraq-Kuwait&5187&1.70&Iran-Iraq&989&2.40&UK-USA&184&2.39\\
SA-USA&4468&1.47&Israel-USA&935&2.27&France-UN&171&2.22\\
Israel-Palestina&4466&1.47&Iraq-Israel&851&2.07&Russia-USA&170&2.20\\
UN-USA&4209&1.38&Iraq-SA&796&1.93&Iraq-Turkey&136&1.76\\
\hline
\hline
\multicolumn{3}{c|}{Apr79-FGW} & \multicolumn{3}{c|}{FGW-Mar99} &  \multicolumn{3}{c}{Iran-Iraq War} \\
\hline
\multicolumn{3}{c|}{Edges 130990} & \multicolumn{3}{c|}{Edges 132230} & \multicolumn{3}{c}{Edges 95189} \\
\hline
Edge  & Degree & \% & Edge & Degree & \% &Edge & Degree & \% \\
\hline
Iran-Iraq&16015&12.23&Iraq-USA&11647&8.81&Iran-Iraq&14470&15.20 \\
Iran-USA&9928&7.58&Iraq-UN&10605&8.02&Iran-USA&6456&6.78 \\
Israel-USA&2714&2.07&Israel-Lebanon&4471&3.38&Israel-USA&2124&2.23 \\
Iran-UN&2105&1.61&Iran-USA&3797&2.87&Iran-UN&1402&1.47 \\
Israel-Lebanon&1981&1.51&UN-USA&2575&1.95&Israel-Lebanon&1391&1.46 \\
Israel-Palestina&1932&1.47&Iraq-Kuwait&2371&1.79&Lebanon-USA&1327&1.39 \\
Lebanon-USA&1722&1.31&UK-USA&2206&1.67&SA-USA&1271&1.34 \\
Lebanon-Syria&1591&1.21&Israel-USA&2154&1.63&Israel-Palestina&1247&1.31 \\
SA-USA&1554&1.19&Israel-Palestina&2144&1.62&France-Iran&1193&1.25 \\
France-Iran&1418&1.08&Iran-Iraq&2117&1.60&Lebanon-Syria&999&1.05 \\
\end{tabular}
\end{center}
\end{table}

\begin{figure}[!t]
\centering
\includegraphics[width=1.0\textwidth]{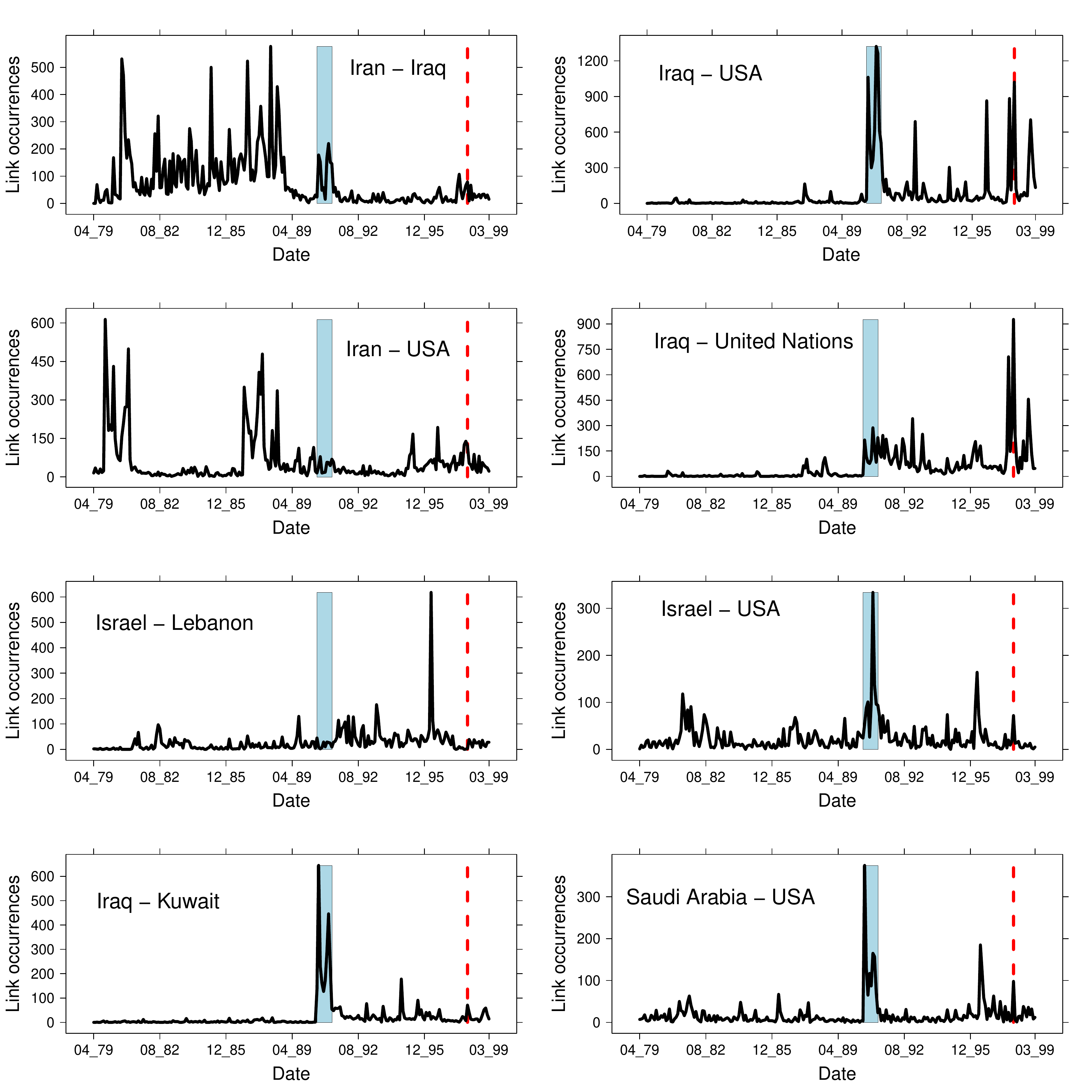}
\caption{Occurrences along time of the top-8 most frequent links. The blue area marks FGW, while the red dashed line indicates IDC in February 98.}
\label{fig:edges}
\end{figure}

In Fig.~\ref{fig:edges} we display the dynamics of the occurrence along time of the top edges, showing their different trends during the diverse events.
It is interesting to note how two top links, Iran-Iraq and Iran-USA are preponderant from 1979 to 1989, \textit{i.e.}, throughout the whole Iran-Iraq War, while they go decaying quickly afterwards, with a minor spike for FGW.
Complementarily, two other major links Iraq-USA and Iraq-United Nations have the opposite trend, remaining almost uninfluential until FGW and growing later on, with a noticeable spike for IDC; moreover, Iraq-United Nations does not show any trend change for FGW, while Iraq-USA does.
The Iraq-Kuwait link has a very limited dynamics, with the unique important spike for FGW. 
Very similar are also the Saudi Arabia-USA and the Israel-USA links, showing an additional lower spike in correspondence of the raise of the terroristic actions between 1995-1996.
This last event is crucial in the Israel-Lebanon relations, where it has the largest effect; FGW, instead, has almost no impact here.

\begin{figure}[!t]
\centering
\includegraphics[width=1.0\textwidth]{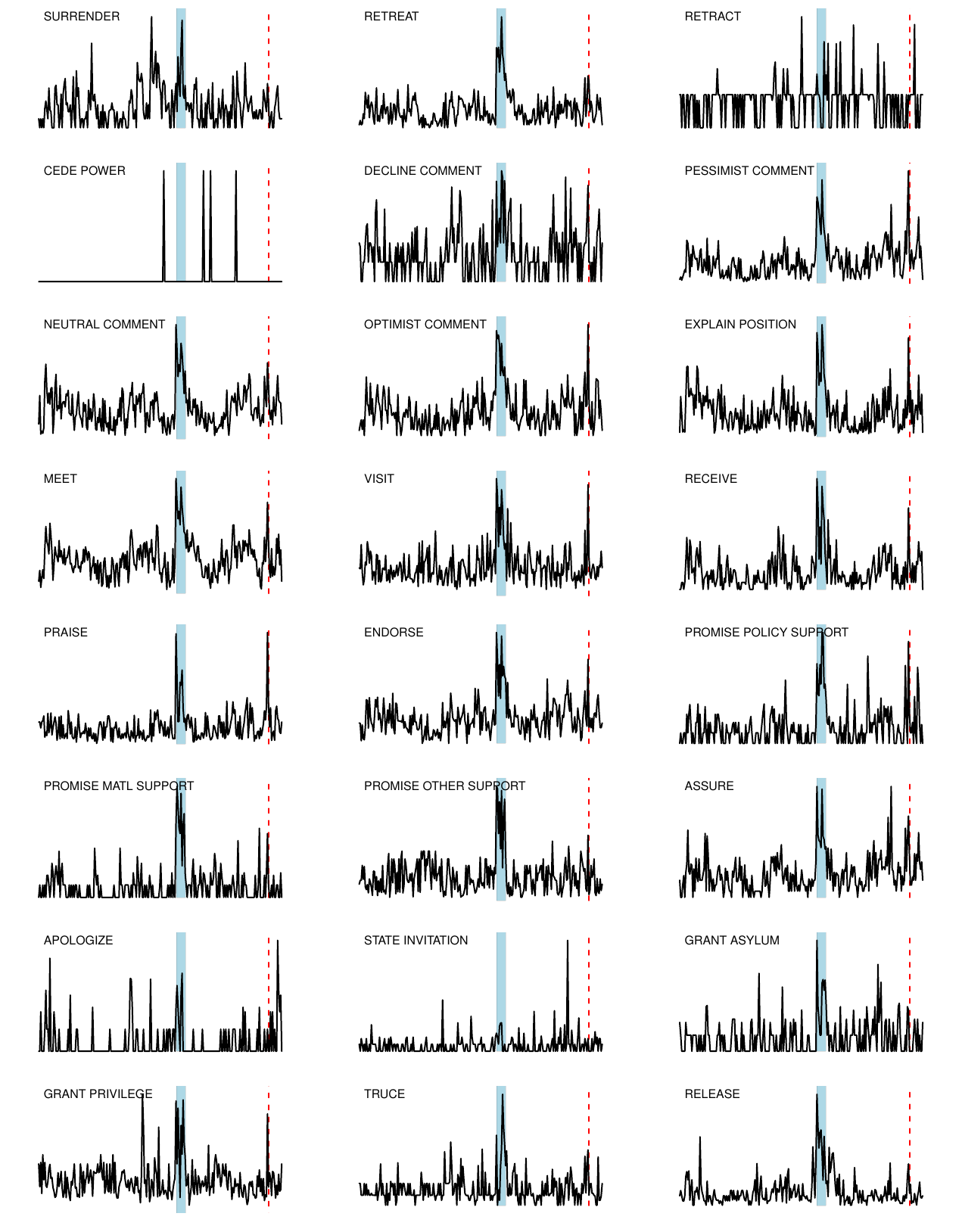}
\caption{Curves of indicator $D1$ for the 24 layers $L_i(t)$, for $i=1,\ldots,24$: the blue area marks FGW, while the red dashed line indicates IDC in February 98. For each curve, the corresponding World Event/Interaction Survey category is indicated in the top left corner.}
\label{fig:d1}
\end{figure}

\begin{figure}[!t]
\centering
\includegraphics[width=1.0\textwidth]{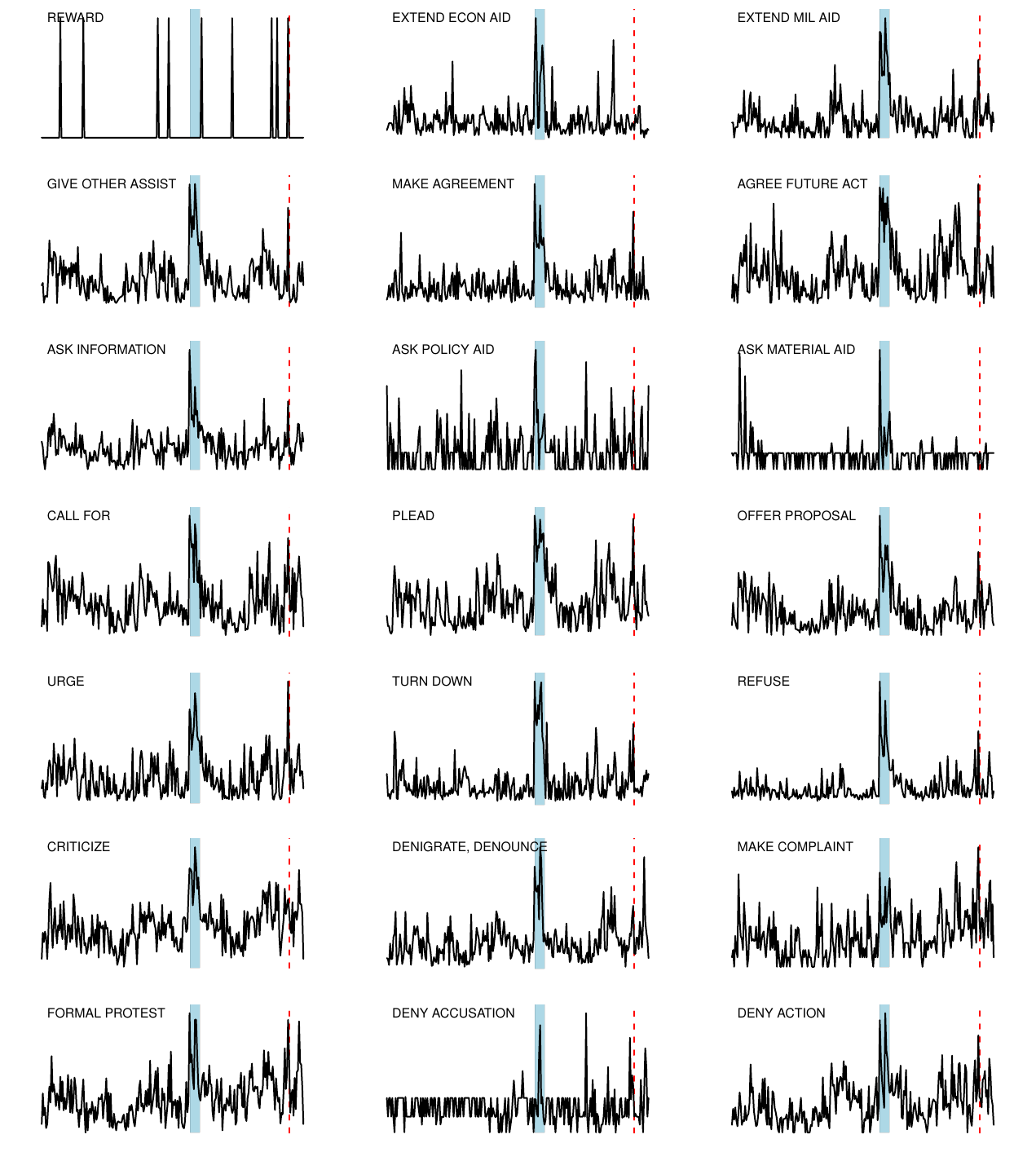}
\caption{Curves of indicator $D1$ for the 21 layers $L_i(t)$, for $i=25,\ldots,45$: the blue area marks FGW, while the red dashed line indicates IDC in February 98. For each curve, the corresponding World Event/Interaction Survey category is indicated in the top left corner.}
\label{fig:d2}
\end{figure}

\begin{figure}[!t]
\centering
\includegraphics[width=1.0\textwidth]{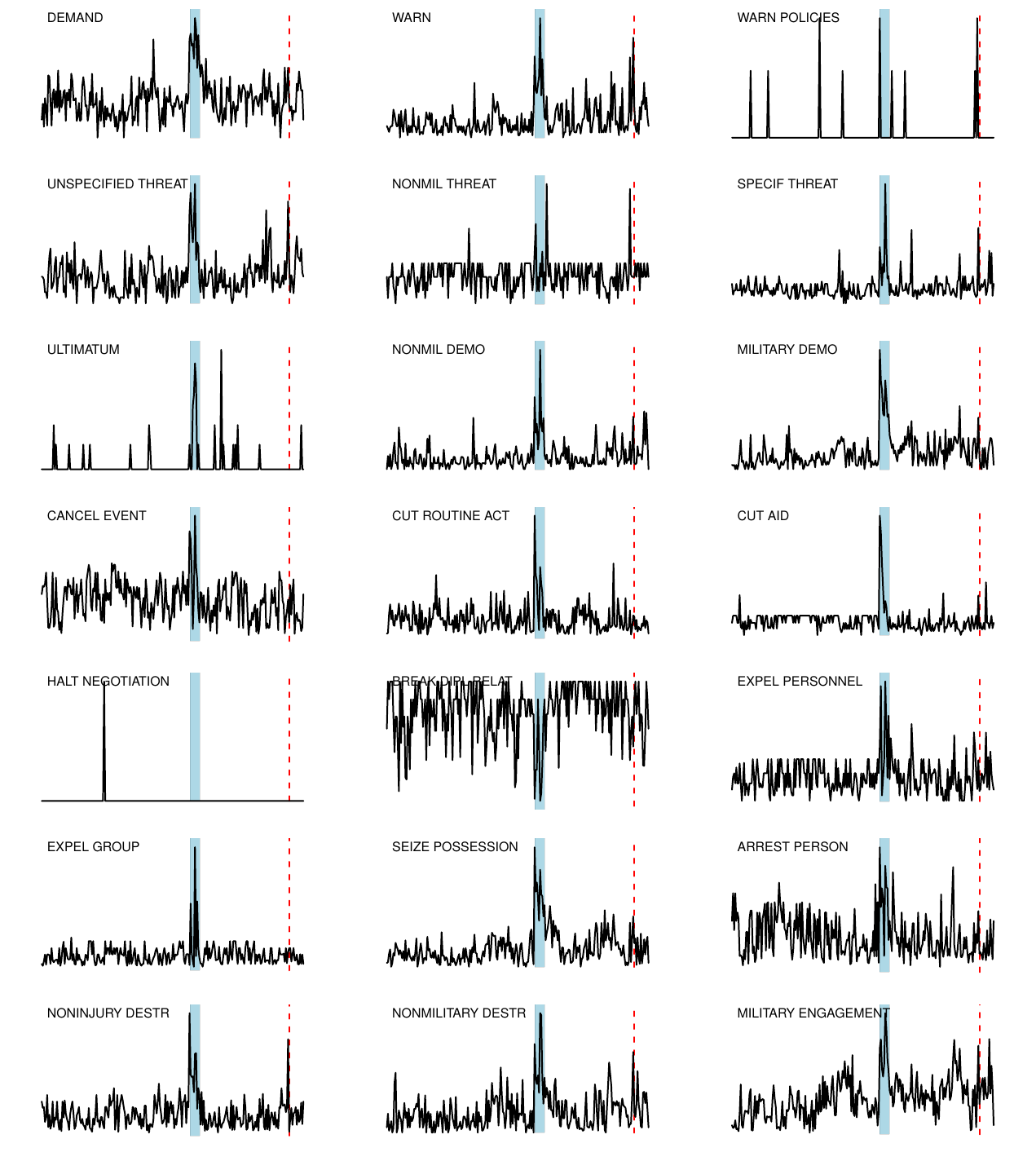}
\caption{Curves of indicator $D1$ for the 21 layers $L_i(t)$, for $i=46,\ldots,66$: the blue area marks FGW, while the red dashed line indicates IDC in February 98. For each curve, the corresponding World Event/Interaction Survey category is indicated in the top left corner.}
\label{fig:d3}
\end{figure}

\begin{figure}[!t]
\centering
\includegraphics[width=1.0\textwidth]{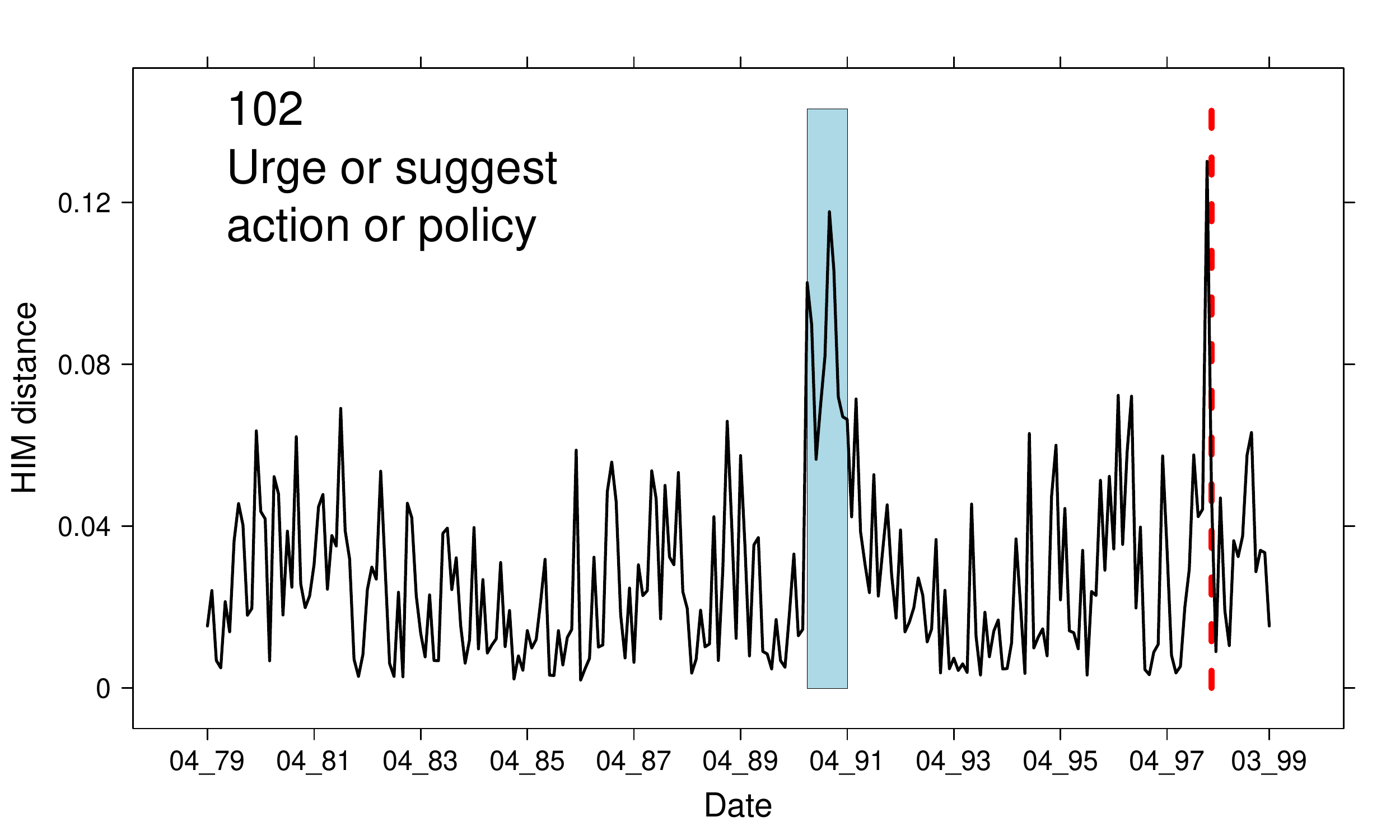}
\caption{$D_1$ time series for the layer 37, corresponding to WEIS code 102 (``Urge or suggest action or policy''). The period corresponding to FGW is marked by the blue background, while the red dashed line indicates IDC in February 1998.}
\label{fig:urge}
\end{figure}

\paragraph{$D*$ indicators analysis}
The two main events FGW and IDC generate sudden changes in the D1 time series for most of the layers: an example is given in Fig.~\ref{fig:urge} for the layer 37, corresponding to WEIS code 102 (``Urge or suggest action or policy''), where we highlight FGW by a blue background, and IDC by a red dashed line.
The complete panel of the $D_1$ curves for all the 66 layers is shown in Fig.~\ref{fig:d1}-\ref{fig:d3}: most of the layers show a decise change in trend in correspondence of the two main events, although some of the layers display a different behaviour (\textit{e.g.}, layer 59, ``Break diplomatic relations''), sometimes due to the paucity of data (\textit{e.g.}, ``Halt negotiations'' or ``Reward'').
Note that many other spikes exist in many layers, corresponding to different geopolitical events occurring throughout the considered timespan.

\begin{figure}[!t]
\centering
\includegraphics[width=1.0\textwidth]{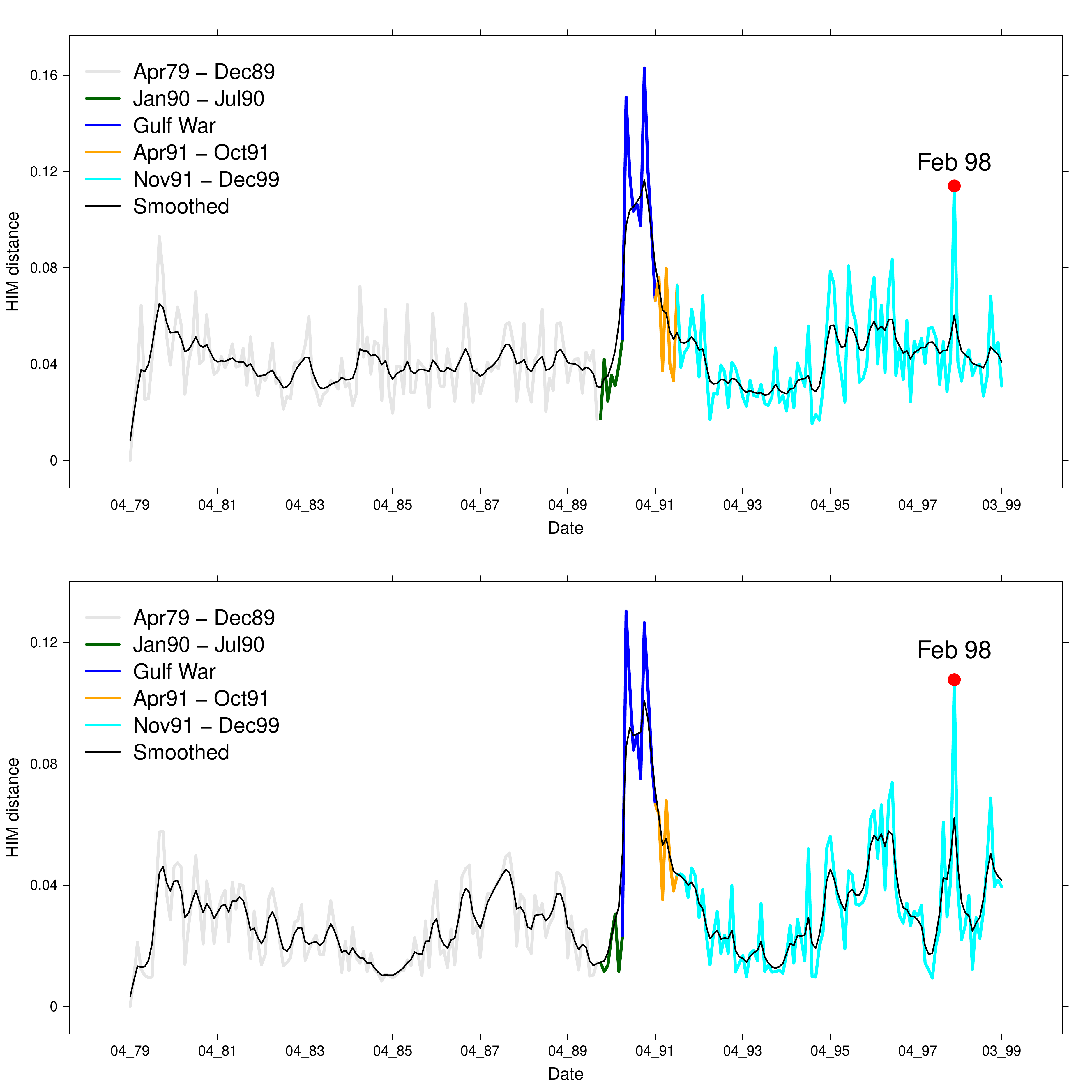}
\caption{Time evolution of a global view of the (monthly) Gulf Dataset. (top) $D_2$ dynamics of the collapsed projections $\{\mathcal{CN}(t)\}^{240}_{t=1}$ and (bottom) $D_3$ dynamics of the metric projections $\{\mathcal{LN}(t)\}^{240}_{t=1}$. For each date, the value on y-axis the is the HIM distance from the first element of the time series. Different colors mark different time periods. The black line represents the fixed-interval smoothing via a state-space model~\cite{durbin01time}.}
\label{fig:gulfts}
\end{figure}

All the information conveyed by the 66 $D_1$ time series can be summarized by using the $D_2$ and $D_3$ indicators displayed in Fig~\ref{fig:gulfts}.
The two curves show a similar trend, with two major spikes corresponding to the FGW and the IDC, neatly emerging in both time series.
Furthermore, both indicators are consistent in showing that the two periods pre- and post-FGW are not part of the FGW spike, implying that in these two periods the structure of the occuring binary geopolitical events is closer to the analogous structure for the ``no-war'' periods.

However, as expected, the indicator $D_2$ includes a lower level of information than $D_3$: this is particularly evident (also for the smoothed curves, in black in the plots) in the periods 85-89 and 95-97, where the dynamics of $D_2$ is much flatter than the dynamics of $D_3$. 
Note that a non trivial dynamics in the two periods 85-89 and 95-97 exist in many layers, as shown in Fig.~\ref{fig:d1}-\ref{fig:d3}, triggered by a number of important events impacting the geopolitical relations: the final part of the Iran-Iraq War (1980-1988), the decline and fall of the Soviet Empire (not direclty related to the Middle East area, but reflecting also there), the dramatic change of the situation of the Middle East conflicts induced by the outbreak of the First Intifada in December 1987~\cite{ragionieri97peace}, and the terrorism excalation (Dhahran, Tel Aviv, and Jerusalem) in Middle East in 95/96 causing a bursting increase in the number of victims just to name the more relevant events.

Thus, this case study, too supports the superiority of $D_3$ as a global indicator to summarize the evolution of a series of multiplex networks.

\begin{figure}[!b]
\centering
\includegraphics[width=0.9\textwidth]{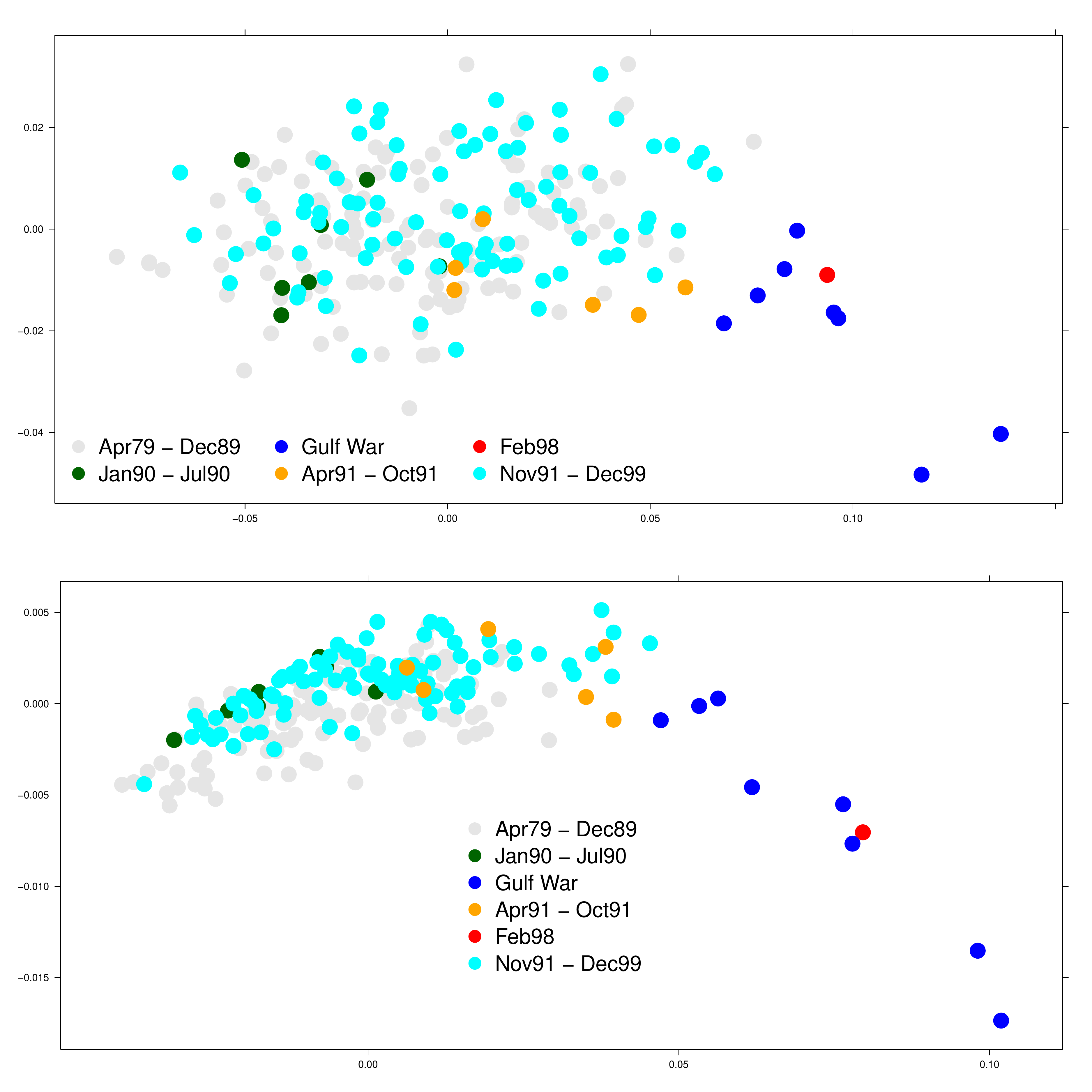}
\caption{Planar multidimensional scaling plot with HIM distance of the collapsed (top) and metric (bottom) projection for the monthly Gulf Dataset. Colors are consistent with those in Fig.~\ref{fig:gulfts}.} 
\label{fig:gulf}
\end{figure}

We also computed all the $\frac{240\cdot 239}{2}$ HIM distances for $D_2$ (respectively, $D_3$) $\left\{\textrm{HIM}(\mathcal{CN}(t_i),\mathcal{CN}(t_j))\right\}_{1\leq i\leq j\leq \tau=240}$ (resp. $\left\{\textrm{HIM}(\mathcal{LN}(t_i),\mathcal{LN}(t_j))\right\}_{1\leq i\leq j\leq \tau=240}$), which are then used to project the 240 networks on a plane through a multidimensional scaling \cite{cox01multidimensional}: the resulting plots are displayed in Fig.~\ref{fig:gulf}. 

Both indicators yield that the months corresponding to FGW (in blue in the plots) are close together and confined in the lower left corner of the plane, showing both a mutual high degree of homogeneity and, at the same time, a relevant difference to the graphs of all other months.
Interestingly, this holds also for the months immediately before and after (in green and orange in the figures) the conflict, which are quite distant from the war months' cloud, as previously observed.
This confirms that, only at the onset of the conflict the diplomatic relations worldwide changed consistently and their structure remained very similar throughout the whole event.

From both the multidimensional scaling plots in Fig.~\ref{fig:gulf} it is clear that the both the $\mathcal{CN}$ and $\mathcal{LN}$ networks for the FGW months can be easily discriminated from all other nets.
However, from the MDS projections it is not evident whether the months Apr 1979 - Dec 1989 (in grey) could be separated from the Nov 91 - Dec 99 months.
By using a Support Vector Machine classifier with the HIM kernel~\cite{jurman15him,jurman14him} (with $\gamma=172.9$ for $\mathcal{LN}$ and $\gamma=110$ for $\mathcal{CN}$), a 5-fold CV classification gives as best result the accuracy 81.2\% for $\mathcal{LN}$ ($C=10^3$) and 73.3\% for $\mathcal{CN}$ ($C=10^4$).
Thus, in both cases, machine learning provides a good separation between the networks belonging to the two periods.

\begin{figure}[!t]
\centering
\includegraphics[width=1.0\textwidth]{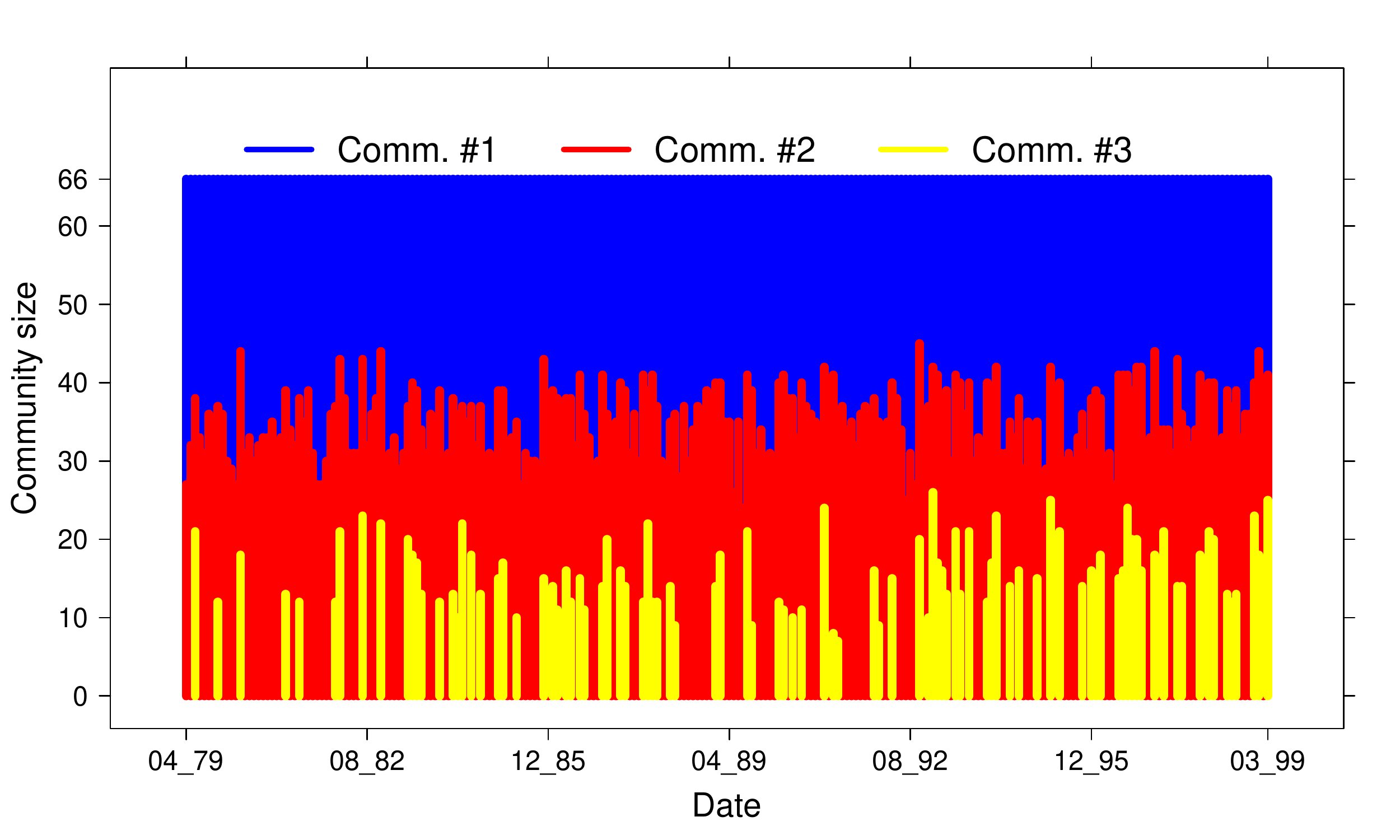}
\caption{Dimension of the three communities indentified by the Louvain algorithm in $\mathcal{LN}$ along the 240 months.}
\label{fig:barplot}
\end{figure}

\begin{figure}[!t]
\centering
\includegraphics[width=1.0\textwidth]{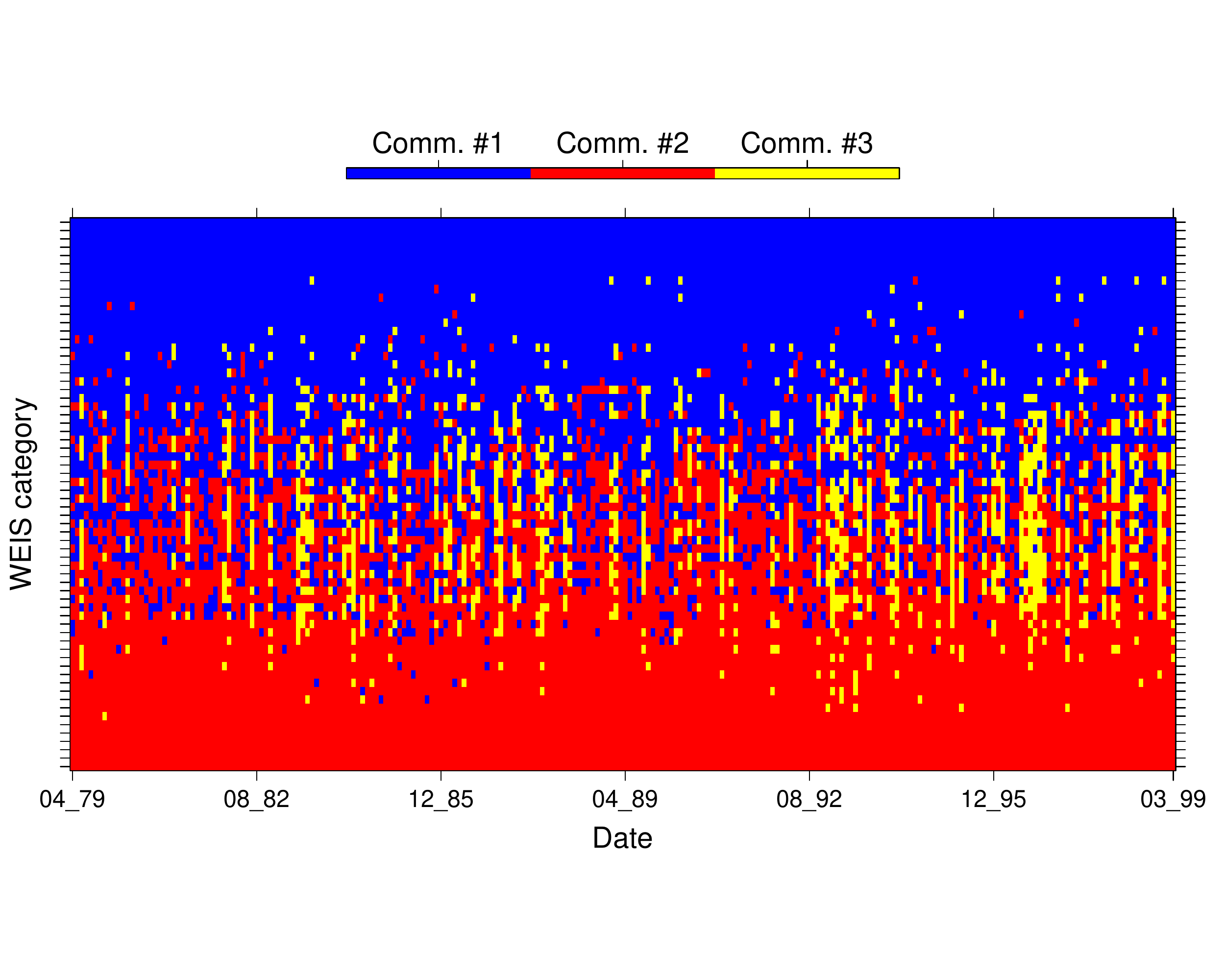}
\caption{Community evolution along time for each of the 66 WEIS categories, ranked by community distribution.}
\label{fig:levelplot}
\end{figure}

\paragraph{Community structure of $\mathcal{LN}$}
We conclude by analyzing the dynamics of the mesostructure of the layer network $\mathcal{LN}$ as extracted by the Louvain community detection algorithm~\cite{blondel08fast,lancichinetti09community,aynaud10optimisation,lancichinetti12consensus}. 
For any temporal step, the Louvain algorithm clusters the 66 nodes (WEIS categories) of $\mathcal{LN}$ into two or three communities, whose dimension along time is shown in Fig.~\ref{fig:barplot}.
In Fig.~\ref{fig:levelplot} we show, for each date, which community each category (on the rows) belongs to; WEIS categories are ranked according to their community distribution, \textit{i.e.}, decreasing number of presences in Comm. \#1 and increasing for Comm. \#2.
Thus in top rows we have the categories lying in Comm. \#1 during all the 240 months (layers 7,10,11,28,34,40), while bottom rows are reserved to the categories always belonging to Comm. \#2 (3,4,19,25,48,52,58): their description in terms of WEIS categories is shown in Tab.~\ref{tab:constant}, while the full community distribution is reported in Tab.~\ref{tab:distribution}.
Focussing on the categories that are consistently lying in a given community throughout all 240 months, some of them are semantically similar: for instance, consult, assistance, action request in community \#1 while two distinct groups emerge in community \#2, namely admit wrongdoing, cede power, apologize, reward on one side and warn of policies, sanction threats and halt negotiations characterizing the second group. 
However, it is interesting the constant presence of the category charge/criticize/blame/disapprove in community \#1.
Many layers sharing the same (or similar) WEIS second level category (Yield, Comment, Consult, etc.) are quite close in the community distribution ranked list, with a general escalating trend proceedings from help request (or other more neutral actions) to more severe situations growing together with the community distribution rank.

\begin{table}[ht]
\centering
\caption{Community distribution for the 66 WEIS categories: for each layer, we report the number of occurences in the three detected communities.}
\label{tab:distribution}
\begin{tabular}{rrrrrrrrrrr}
  \hline
Rank & Layer & \#1 & \#2 & \#3 & & Rank & Layer & \#1 & \#2 & \#3 \\
  \hline
 1 & 7 & 240 & 0 & 0   &$\qquad$& 34 & 8 & 86 & 108 & 46 \\ 
 2 & 10 & 240 & 0 & 0  &$\qquad$& 35 & 17 & 86 & 106 & 48 \\ 
 3 & 11 & 240 & 0 & 0  &$\qquad$& 36 & 24 & 71 & 130 & 39 \\ 
 4 & 28 & 240 & 0 & 0  &$\qquad$& 37 & 12 & 58 & 131 & 51 \\ 
 5 & 34 & 240 & 0 & 0  &$\qquad$& 38 & 47 & 56 & 130 & 54 \\ 
 6 & 40 & 240 & 0 & 0  &$\qquad$& 39 & 27 & 54 & 140 & 46 \\ 
 7 & 66 & 239 & 0 & 1  &$\qquad$& 40 & 26 & 51 & 154 & 35 \\ 
 8 & 9 & 236 & 0 & 4   &$\qquad$& 41 & 18 & 45 & 153 & 42 \\ 
 9 & 37 & 234 & 2 & 4  &$\qquad$& 42 & 42 & 31 & 170 & 39 \\ 
 10 & 38 & 234 & 1 & 5 &$\qquad$& 43 & 53 & 27 & 174 & 39 \\ 
 11 & 29 & 233 & 2 & 5 &$\qquad$& 44 & 61 & 26 & 179 & 35 \\ 
 12 & 30 & 232 & 2 & 6 &$\qquad$& 45 & 51 & 21 & 182 & 37 \\ 
 13 & 6 & 230 & 2 & 8  &$\qquad$& 46 & 1 & 17 & 204 & 19 \\ 
 14 & 35 & 230 & 0 & 10&$\qquad$& 47 & 60 & 15 & 204 & 21 \\ 
 15 & 41 & 222 & 3 & 15&$\qquad$& 48 & 57 & 14 & 214 & 12 \\ 
 16 & 39 & 220 & 7 & 13&$\qquad$& 49 & 50 & 9 & 221 & 10 \\ 
 17 & 43 & 208 & 12 & 20&$\qquad$& 50 & 59 & 7 & 225 & 8 \\ 
 18 & 14 & 196 & 18 & 26&$\qquad$& 51 & 23 & 5 & 219 & 16 \\ 
 19 & 62 & 175 & 29 & 36&$\qquad$& 52 & 32 & 5 & 231 & 4 \\ 
 20 & 63 & 170 & 32 & 38&$\qquad$& 53 & 15 & 4 & 232 & 4 \\ 
 21 & 2 & 164 & 28 & 48 &$\qquad$& 54 & 16 & 2 & 236 & 2 \\ 
 22 & 56 & 150 & 51 & 39&$\qquad$& 55 & 20 & 2 & 236 & 2 \\ 
 23 & 46 & 148 & 49 & 43&$\qquad$& 56 & 33 & 2 & 237 & 1 \\ 
 24 & 36 & 146 & 48 & 46&$\qquad$& 57 & 5 & 1 & 235 & 4 \\ 
 25 & 55 & 136 & 46 & 58&$\qquad$& 58 & 21 & 1 & 238 & 1 \\ 
 26 & 45 & 122 & 64 & 54&$\qquad$& 59 & 44 & 1 & 231 & 8 \\ 
 27 & 64 & 121 & 71 & 48&$\qquad$& 60 & 3 & 0 & 240 & 0 \\ 
 28 & 65 & 119 & 64 & 57&$\qquad$& 61 & 4 & 0 & 240 & 0 \\ 
 29 & 31 & 116 & 64 & 60&$\qquad$& 62 & 19 & 0 & 240 & 0 \\ 
 30 & 22 & 110 & 79 & 51&$\qquad$& 63 & 25 & 0 & 240 & 0 \\ 
 31 & 54 & 109 & 86 & 45&$\qquad$& 64 & 48 & 0 & 240 & 0 \\ 
 32 & 13 & 97 & 91 & 52 &$\qquad$& 65 & 52 & 0 & 240 & 0 \\ 
 33 & 49 & 97 & 80 & 63 &$\qquad$& 66 & 58 & 0 & 240 & 0 \\ 
   \hline
\end{tabular}
\end{table}

\begin{table}[!ht]
\centering
\caption{The 13 layers not swapping community across all 240 timepoints.}
\label{tab:constant}
\begin{tabular}{rrl}
\hline
\multicolumn{3}{c}{Community \#1} \\
Layer & WEIS code & WEIS category  \\
\hline
7  & 023 & [Comment] Comment on situation – neutral  \\
10 & 031 & [Consult] Meet with at neutral site, or send note \\ 
11 & 032 & [Consult] Consult \& Visit; go to  \\
28 & 073 & [Reward] Give other assistance \\
34 & 094 & [Request] Request action; call for \\
40 & 121 & [Accuse] Charge; criticize; blame; disapprove \\
\hline
\multicolumn{3}{c}{Community \#2}\\
Layer & WEIS code & WEIS category \\
\hline
3  & 013 & [Yield] Admit wrongdoing; retract statement \\
4  & 015 & [Yield] Cede power \\                                                            
19 & 061 & [Grant] Express regret; apologize \\                                             
25 & 070 & [Reward] Reward \\                                                               
48 & 161 & [Warn] Warn of policies \\
52 & 174 & [Threaten] Ultimatum; threat with negative sanctions and time limit specified \\ 
58 & 194 & [Reduce Relations] Halt negotiations \\
\hline
\end{tabular}
\end{table}

\begin{figure}[!t]
\centering
\includegraphics[width=0.95\textwidth]{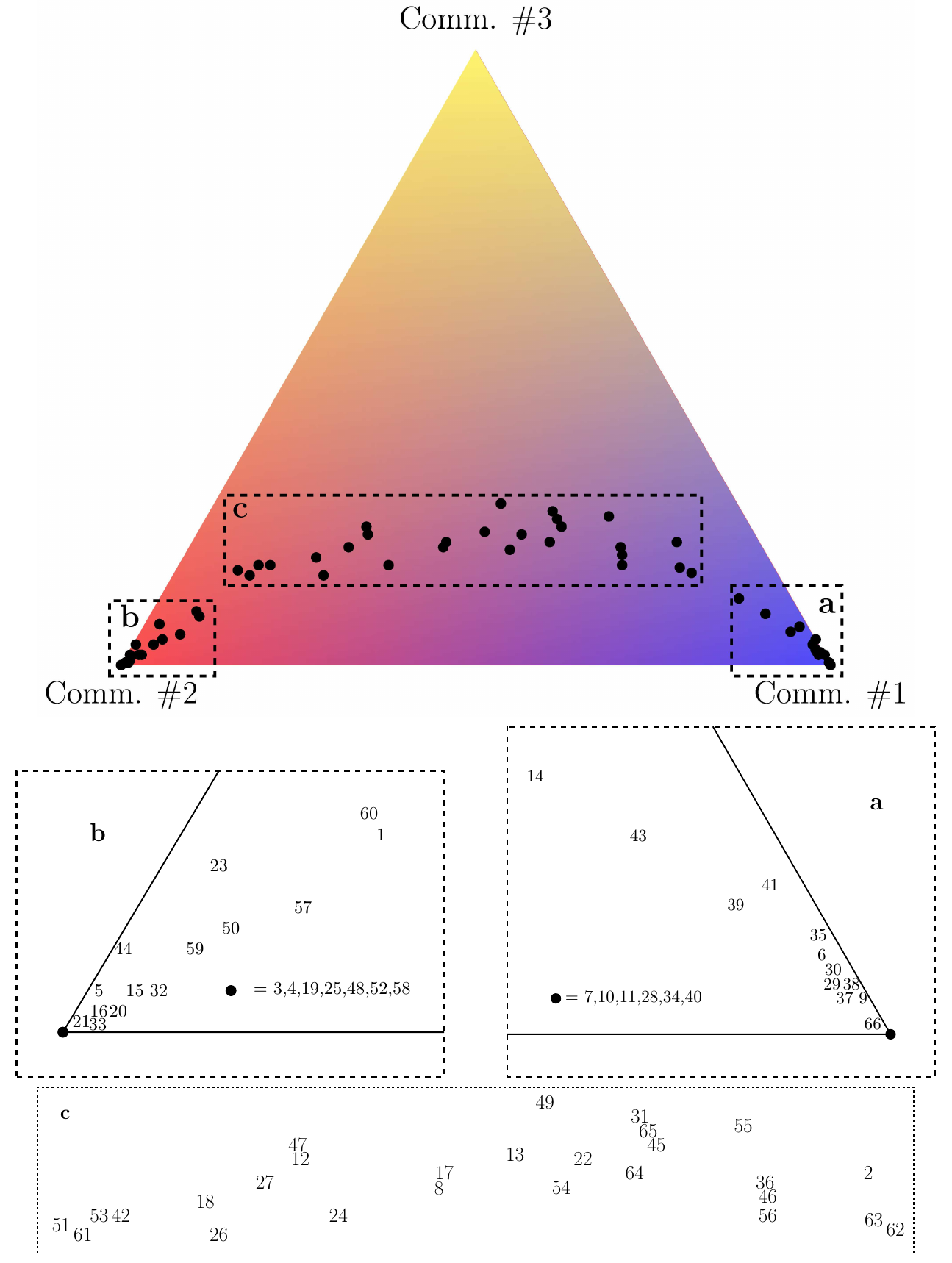}
\caption{Triangleplot projection of the 66 WEIS categories defined by their community distribution.}
\label{fig:triangle}
\end{figure}

\section{Conclusion}
\label{sec:conclusion}
We introduced here a novel approach for the longitudinal analysis of a time series of multiplex networks, defined by mean of a metric transformation conveying the information carried by all layers into a single network for each timestamp, with the original layers as nodes.
The transformation is induced by the Hamming-Ipsen-Mikhailov distance between graph sharing the same nodes, and it preserves the key events encoded into each instance of the multiplex network time series, making it more efficient than the collapsing of all layers into one collecting all edges for detecting inportant fluctuations in the original network's dynamics.
Moreover, a community detection analysis on the obtained network can help shading light on the relations between the original layers throughout the whole time span.

\section*{References}
\bibliography{jurman16metric}
\end{document}